\documentclass[12pt,twoside]{article}
\usepackage[T1]{fontenc}
\usepackage[latin1]{inputenc}
\usepackage{times}
\usepackage{graphicx}
\usepackage{a4wide}
\usepackage{listings}
 \usepackage{amsmath}
\usepackage{amssymb}

\begin{document}

\section*{Radiation Tolerance of CMOS Monolithic Active Pixel Sensors with Self-Biased Pixels
  \footnote{Presented at 11th European Symposium on Semiconductor Detectors, 
    June 7 - 11, 2009 at Wildbad Kreuth
}}

\thispagestyle{empty}

\begin{raggedright}

\markboth{M. Deveaux {\it et al.}}
{Radiation Tolerance of CMOS Monolithic Active Pixel Sensors with Self-Biased Pixels}

M.~Deveaux$^1$,
S.~Amar-Youcef$^1$,
A.~Besson$^2$,
G.~Claus$^2$,
C.~Colledani$^2$,
M.~Dorokhov$^2$,
C.~Dritsa$^{1,2,3}$,
W.~Dulinski$^2$,
I.~Fr\"ohlich$^1$,
M.~Goffe$^2$,
D.~Grandjean$^2$,
S.~Heini$^2$,
A.~Himmi$^2$,
C.~Hu$^2$,
K.~Jaaskelainen$^2$,
C.~M\"untz$^1$,
A.~Shabetai$^2$,
J.~Stroth$^{1,3}$,
M.~Szelezniak$^4$,
I.~Valin$^2$ and
M.~Winter$^2$
\vspace{1cm}\\
\hspace{-0.4cm}\makebox[0.3cm][r]{$^{1}$}
Institut f\"{u}r Kernphysik, Johann Wolfgang Goethe-Universit\"{a}t, 60438 ~Frankfurt, Germany\\
\hspace{-0.4cm}\makebox[0.3cm][r]{$^{2}$}
Institut Pluridisciplinaire Hubert Curien (IPHC),
67037 Strasbourg, France\\
\hspace{-0.4cm}\makebox[0.3cm][r]{$^{3}$}
Gesellschaft f\"{u}r Schwerionenforschung mbH, 64291~Darmstadt, Germany\\
\hspace{-0.4cm}\makebox[0.3cm][r]{$^{4}$}
Lawrence Berkeley National Laboratory, Berceley CA 94720, U.S.A.
\end{raggedright}
\vspace{1cm}
\begin{center}
\textbf{Abstract}
\vspace{0.5cm}\\
\begin{minipage}[c]{0.9\columnwidth}
CMOS Monolithic Active Pixel Sensors (MAPS) are proposed as a
technology for various vertex detectors in nuclear and particle
physics. We discuss the mechanisms of ionizing radiation damage on
MAPS hosting the the dead time free, so-called self bias
pixel. Moreover, we discuss radiation hardened sensor designs which
allow operating detectors after exposing them to irradiation doses
above \mbox{1 Mrad}.
\end{minipage}

\end{center}

\section{Introduction}

The ability of Monolithic Active Pixel Sensors (MAPS) to provide
charged particle tracking has been demonstrated with several
MIMOSA\footnote{Standing for Minimum Ionizing MOS Active pixel
  sensor.}  prototypes \cite{MAPS-TestPaper,MIMOSA4Paper}.  The key
element of such sensors is the use of an n-well/p-epitaxial diode to
collect, through thermal diffusion, the charge generated by the
penetrating particles in the thin epitaxial layer located underneath
the readout electronics. Tests performed with \mbox{$120~$GeV/c} pion
beams at CERN proved excellent detection performances.  A single point
resolution of \mbox{$1 - 2 ~\mu$m} and a detection efficiency close to
100\%, resulting from a signal-to-noise ratio (S/N) in excess of 30,
were routinely observed with various MAPS designs featuring up to
10$^6$ pixels on active areas as large as \mbox{$4~$cm$^2$}.

\label{sectionChipOperationAllPixels}
\begin{figure}[]
  \begin{minipage} [c]{5cm}  
    \includegraphics[width=0.8\columnwidth]{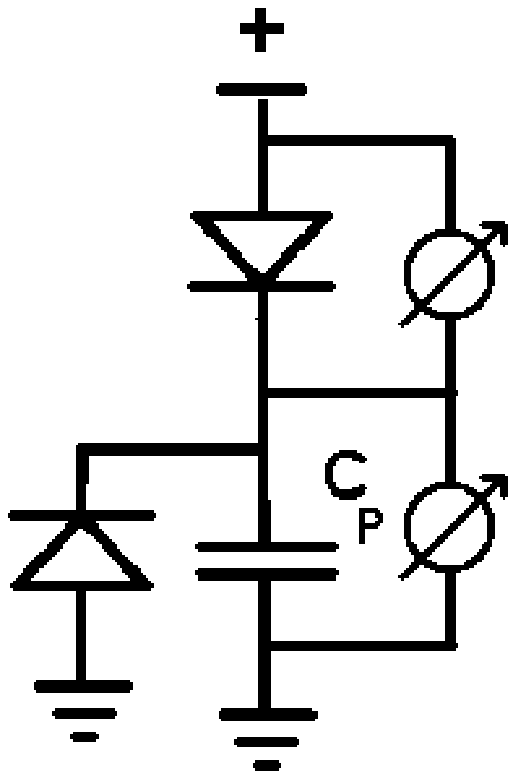}
  \end{minipage}    
  \begin{minipage} [c]{8cm}  \hspace{2cm}
    \includegraphics[width=0.8\columnwidth]{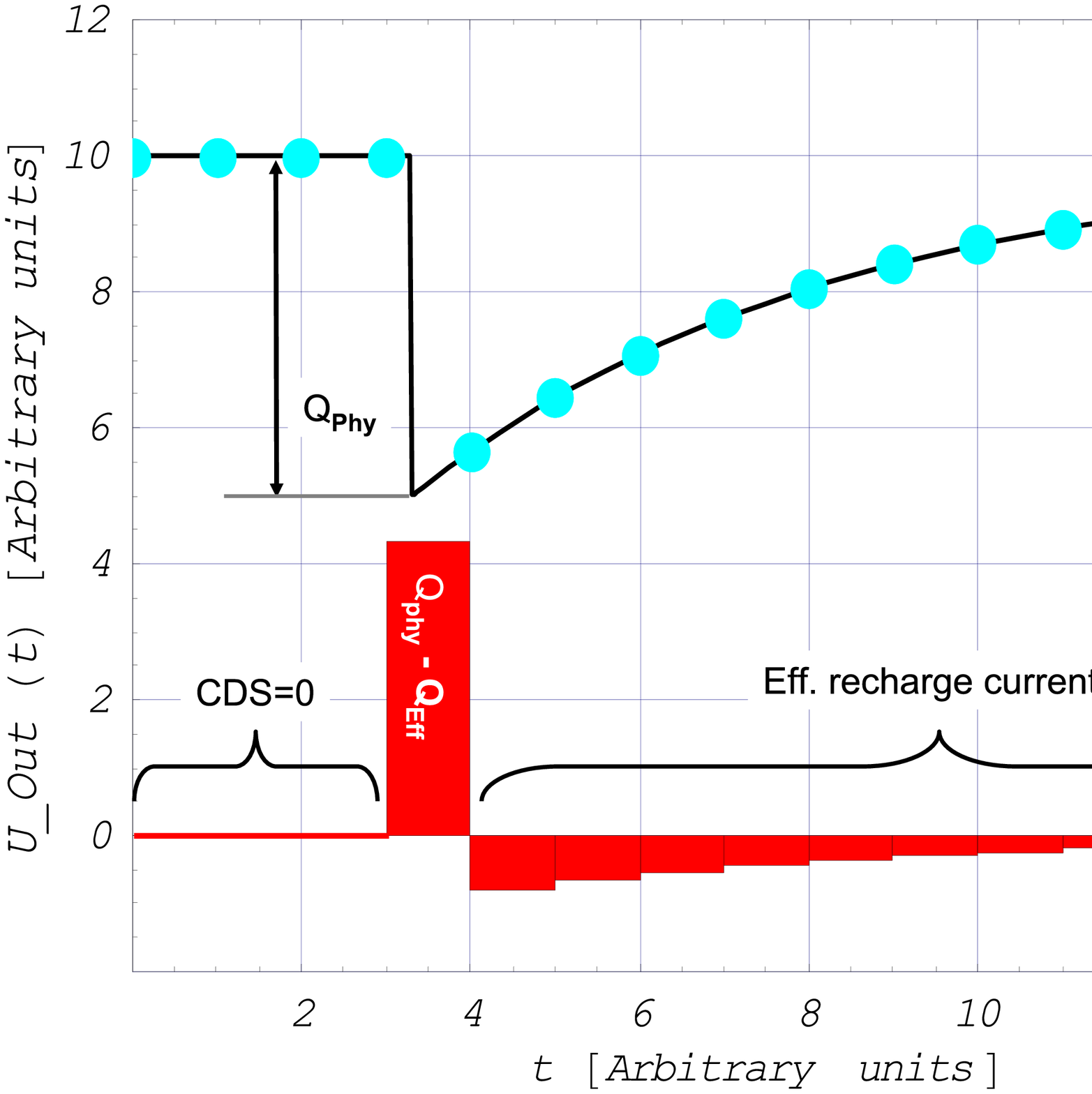}        
  \end{minipage}   \vspace{0.5cm}
  \caption[]{Simplified diagram of the SB-pixel (left) and the
    response of this pixel to a hit injecting a signal charge of
    $Q_S=Q_{Phy}$ (right). Colors online.}
  \label{fig:SBPixelReadout}  
\end{figure}

MAPS have been proposed as sensor technology for the vertex detectors
of the International Linear Collider (ILC), the STAR Heavy Flavor
Tracker and the Compressed Baryonic Matter (CBM) experiment. The
expected integrated radiation doses of in those application ranges
from several $10^{10}~$n$_{\rm eq}/$cm$^2$ and few 100 krad (ILC) to
up to \mbox{$\sim 10^{15}~$n$_{\rm eq}/$cm$^2$} and $\sim 100~\rm
Mrad$ (CBM).  The question whether MAPS for charged particle tracking
can stand these doses triggered a joint research program of the
IPHC/Strasbourg, the University of Frankfurt and GSI/Darmstadt.

In this work, we will discuss the impact of ionizing radiation on MAPS
pixels with a so-called self-biased on-pixel amplifier (SB-pixel)
\cite{MIMOSA4Paper}. To do so, we will remind the design and readout
principle of this pixel and show up the observed radiation damage
effect. Hereafter, we will motivate that all effects observed can be
explained as non-trivial consequences of the known radiation induced
increase of leakage currents. Finally, we will discuss a radiation
hardened pixel structure and show selected test results proving the
higher radiation tolerance of this pixel with respect to the standard
pixel design.

\section{The MAPS operation principle}
 
The sensor of MAPS is formed by a sandwich of three differently
P-doped silicon layers, which are the highly doped substrate, the
moderately doped epitaxial layer and a highly doped P-well layer found
in standard CMOS processes. Due to build-in-voltages, free electrons
created by penetrating charged particles are deflected back toward the
epitaxial layer when reaching the interface to the highly doped outer
layers. The diffusing charge carriers are collected from the $\rm \sim
10~\mu m$ thick, non-depleted sensor via a collection diode formed
from N-well implantation penetrating the P-well layer.
 
The collected charge is loaded into the parasitic capacity of the
pixel amplifier displayed in figure
\ref{fig:SBPixelReadout}(left). The corresponding voltage drop in this
capacity is measured by means of a source follower. A high ohmic,
forward biased diode is used clear the signal from the pixel and to
compensate the leakage current of the collection diode. The clearing
process has to be slow with respect to the pixel readout. Therefore,
the signal charge remains for several integration cycles in the
pixel. The use of Correlated Double Sampling (CDS) avoids a double
counting of hits as pixels clearing a signal provide a modest negative
signal after CDS. This readout cycle and the response of the SB-pixel
to a hit is displayed in figure \ref{fig:SBPixelReadout} (right).

\section{Ionizing radiation damage in SB-pixels}
\begin{figure}[tb]
\begin{center}   
\includegraphics[angle=0, width=0.48\columnwidth]{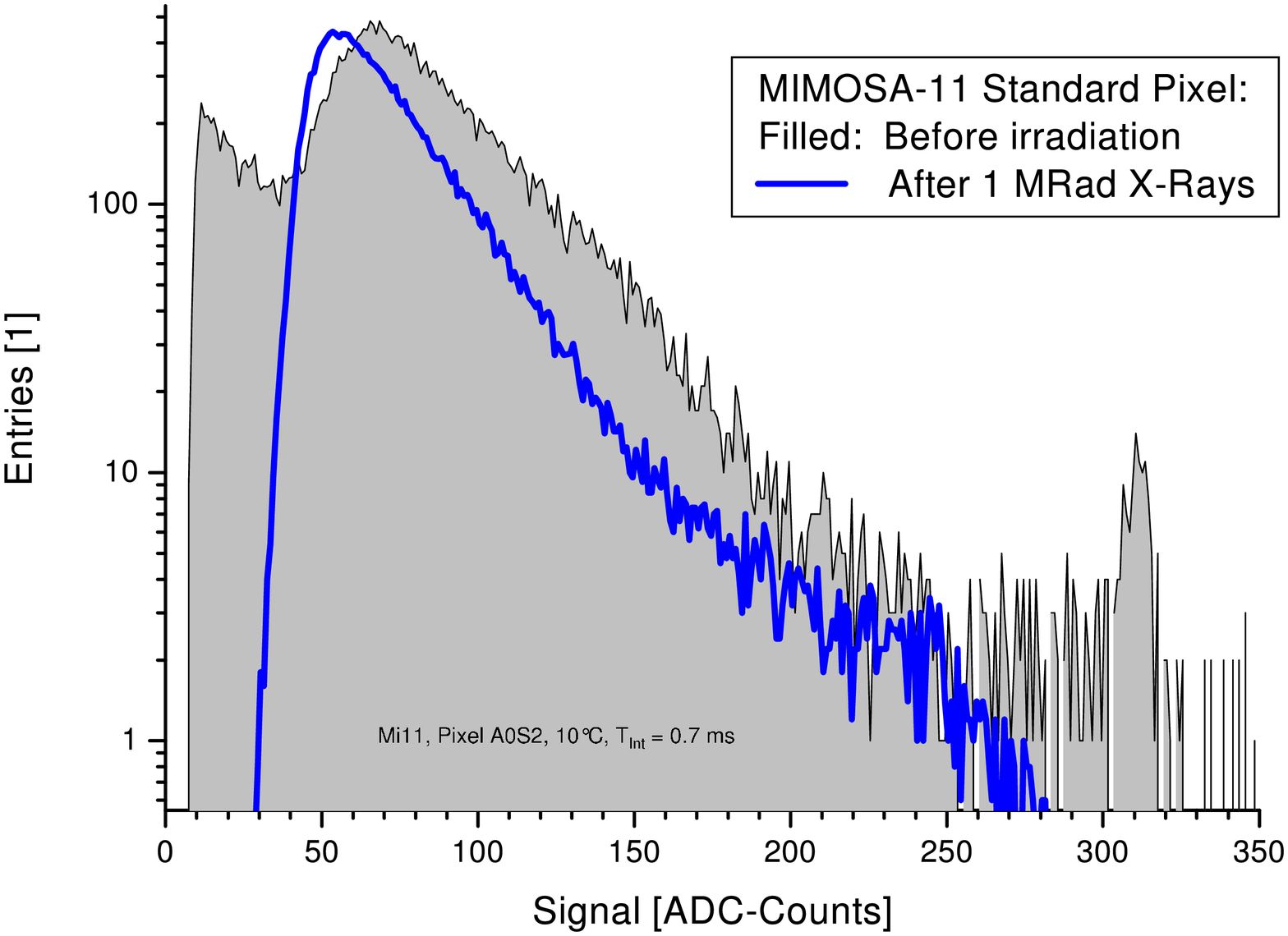}
\includegraphics[angle=0, width=0.48\columnwidth]{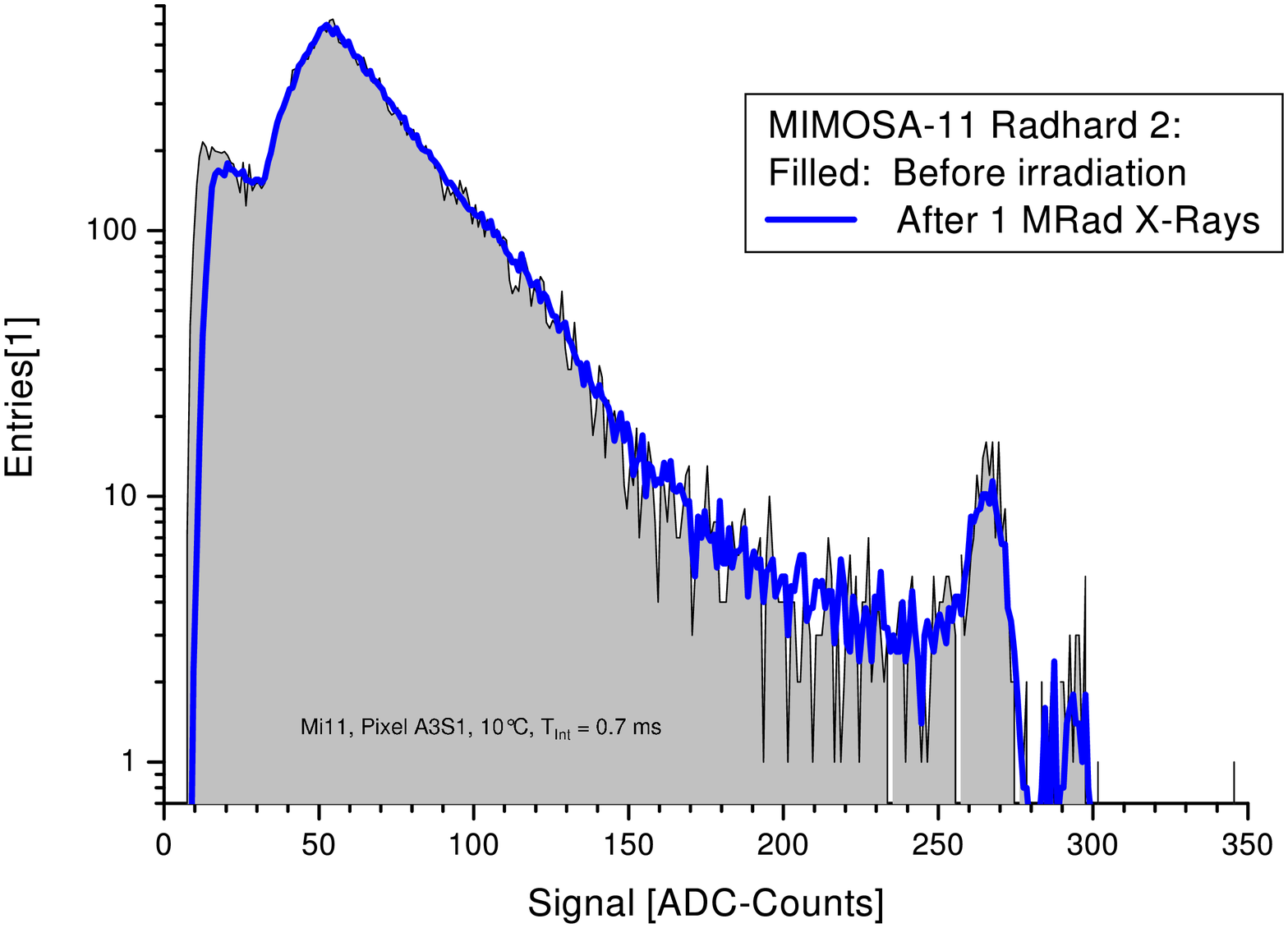}
\end{center}
  \caption[The seed pixel spectrum of MIMOSA-11 pixel A0S2 before and
    after \mbox{$1 \ MRad$}.]  {Response of a standard (left) and a
    radiation hardened (right) SB-pixels to a $^{55}$Fe source. The
    small peak at 320 (left) / 270 (right) ADC-counts corresponds to a
    \mbox{100 \%} charge collection efficiency. The distributions were
    measured at \mbox{$T=10 \rm \ ^{\circ}C$} with an integration time
    of \mbox{$0.7 \ \rm ms$}.}
   \label{Mi11SpectrumA0S2-1MRad}
\end{figure}

\begin{figure}[tbp]
\begin{center} 
 \begin{minipage}[c]{7cm}  
   \includegraphics[angle=0, width=0.9\columnwidth]{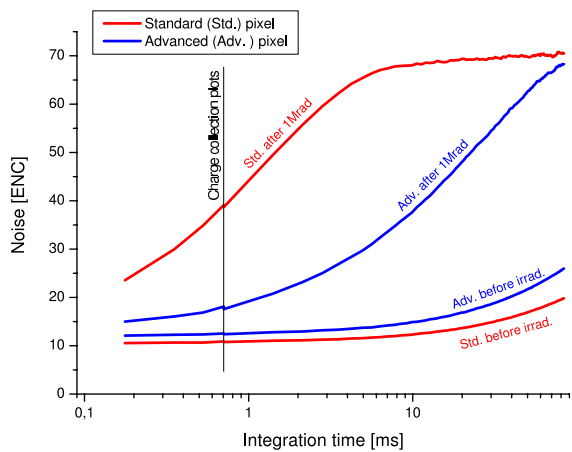}\vspace{1cm}
 \end{minipage}
  \begin{minipage}[c]{6cm}
	\caption[Noise as a function of ionizing doses, temperature
          and the integration on most radiation hard and the standard
          pixel of MIMOSA-11.]  { Noise of the standard (red) and the
          radiation hard (blue) pixel of MIMOSA-11 as function of the
          integration time at a temperature of T=10 $°$C. Colors
          online.}
   \end{minipage}
\end{center}
  
   \label{Mi11Noise2}
\end{figure}

The effects of ionizing radiation on SB-pixels was studied with
several generations of MIMOSA prototypes being manufactured in the
AMS-0.35 $\mu$m and AMS-0.35 $\mu$m OPTO process. The chips were
tested with an $^{55}$Fe-source, irradiated with the X-ray source of
the Institute of Experimental nuclear Physics (IEKP, University of
Karlsruhe) and tested again. For radiation doses of up to 1 Mrad, we
observed two major effects, which could each alleviated by cooling and
reducing the integration time of the sensor: a loss of gain of the
pixel, which smeared out the peak corresponding to an interaction of
the $^{55}$Fe-photon in the small depleted volume of the sensor (see
figure \ref{Mi11SpectrumA0S2-1MRad}, left); and increase of noise (see
figure \ref{Mi11Noise2}). At high temperatures, radiation doses and
integration times, the noise reached a saturation level of \mbox{$\sim
  70$ ENC}, despite (according to all our experiences) the pixel
leakage currents increase further. The saturation stands therefore in
contrast to the expected behavior of shot noise, which should scale
with $Q_{Noise} = e \cdot \sqrt {N_{CollectedElectrons}}$. We consider
the continuous clearing of the pixel as origin for the misfit between
expected and observed behavior of the pixel.

\section{The relation between gain loss and leakage current}

\begin{figure}[tbp]
\begin{center} 
 \begin{minipage}[c]{8cm}  
   \includegraphics[angle=0, width=0.9\columnwidth]{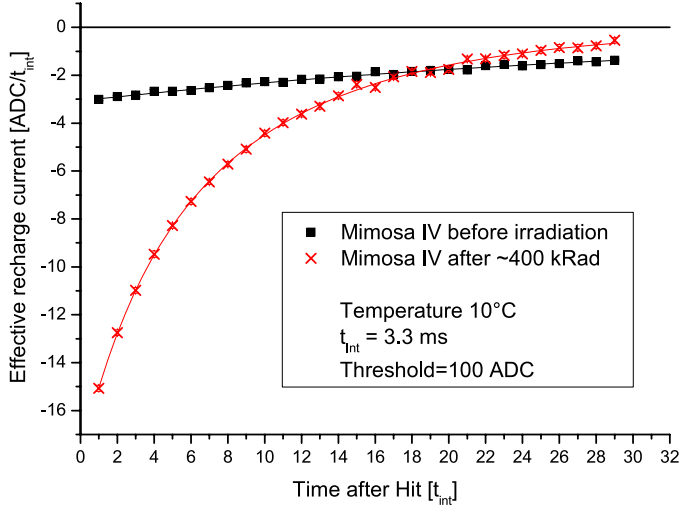}\vspace{1cm}
 \end{minipage}
  \begin{minipage}[c]{5.1cm}
	\caption[] {The effective recharge current $dQ_S/dt$ as
          function of the time after a hit. One ADC-unit corresponds
          to a charge of \mbox{$\sim 11$ e. Colors online.} }
   \end{minipage}
\end{center}
  
   \label{Mi4RechargeCurrent}
\end{figure}

The accelerated clearing of the pixels (see figure
\ref{Mi4RechargeCurrent}) was identified as origin of their effective
gain loss. In irradiated pixels, a substantial part of the signal is
cleared already before it is read out.

Our measurements show, that this clearing process is well described by
an exponential function according to:
\begin{equation}
Q_S(t)= Q_S(t=0) \cdot \exp (- t/\tau)
\label{eqn:PixelResponse}
\end{equation}
In this equation, $Q_S(t)=U_S(t)/C$ is the remaining signal charge in
the pixel capacity $C$ and $\tau$ is the time constant of the
pixel. In order to sense the signal before it is cleared, the
integration time $t_{int}$ of the pixel has fulfill ($t_{int} \ll
\tau$). This is usually assured by the high ohmic design of the
biasing diode.

However, as we will motivate in the following, $\tau$ shrinks with
increasing leakage current of the pixel ($I_L$, see figure
\ref{fig:SBPixelReadout}, left). To show this, we assume that the
potential drop $U_2$ at the biasing diode is split into a contribution
$U_{L}$ and $U_S$ where $U_{L}$ is required to drive the equivalent of
the leakage current $I_L$ through this diode.  The diode fulfills
therefore:
\begin{eqnarray}
I_L&=& I_S \cdot \left ( \exp \left[ \Phi \cdot U_L \right ] -1 \right
) \quad \rm (in ~equilibrium) \label{eqn:gleichgewicht}\\ I_{Eff}(t) +
I_L &=& I_S \cdot \left ( \exp \left[ \Phi \cdot ( U_L + U_S(t))
  \right ] -1 \right ) \quad \rm (with ~signal)
\end{eqnarray}
In those equations, $\Phi= e / (k_B~T)$ and $I_{Eff}(t)= - dQ_S(t)/dt$
is the effective recharge current of the pixel capacity. From the
above equations we derive:
\begin{equation}
I_{Eff}(t)=I_S \cdot \exp[\Phi \cdot U_{L}] \cdot \left (\exp[\Phi \cdot U_S(t)] -1 \right )
\end{equation}
Accounting for equation (\ref{eqn:gleichgewicht}), this is equivalent
to:
\begin{equation}
I_{Eff}(t)= (I_L + I_S) \cdot \left (\exp[\Phi \cdot U_S(t)] -1 \right )
\end{equation}
Using the leading order approximation:
\begin{equation}
\exp[\Phi \cdot U_S(t)] -1 = \Phi \cdot U_S(t) + \mathcal{O}[~\Phi \cdot U_{S}(t)~]^2
\label{eqn:approx}
\end{equation}
 and knowing that $U_S(t)=Q_S(t)/C$ and $I_{Eff}(t)= - dQ_S(t)/dt$,
 this can be translated to:
\begin{equation}
-\frac{d Q_s(t)}{dt} = (I_L + I_S) \cdot \Phi \cdot \frac{Q_S(t)}{C}
\end{equation}
The solution of this differential equation fits equation
(\ref{eqn:PixelResponse}) and thus our empiric observation.  One
identifies:
\begin{equation}
\tau = \frac{C}{\Phi \cdot (I_L + I_S)}
\label{eqn:tauIL}
\end{equation}
From this equation one can learn that the radiation induced increase
of $I_L$ in MAPS \cite{Paper:RESMDD04} may crucially reduce $\tau$,
which explains the observed acceleration of the pixel clearing.

Note that, due to the rather rough approximation made in equation
(\ref{eqn:approx}), equation (\ref{eqn:tauIL}) is not straight
forwardly suited to predict the absolute value or temperature
dependence of $\tau$. This dependence was addressed by
measuring\footnote{This was done by illuminating the chip with an
  Fe$^{55}$-source and observing the clearing of the pixel. Details of
  the measurement protocols and additional results are discussed in
  \cite{Doktorarbeit}.} $\tau$ as a function of temperature with
SB-pixels irradiated with 400 krad X-rays. As shown in figure
\ref{Mi11Tau} (left), $\tau$ becomes by more than one order of
magnitude shorter when increasing the temperature from -10 $^\circ$C
to 40 $^\circ$C, which may be explained by the known strong increase
of $I_L$ with increasing temperature. Cooling is therefore an
effective mean to alleviate ionizing radiation damage in SB-pixels.

\section{Pixels with improved radiation hardness}
\begin{figure}[tbp]
\begin{center} 
 \begin{minipage}[c]{6.5cm}  
 \includegraphics[width=5.5cm]{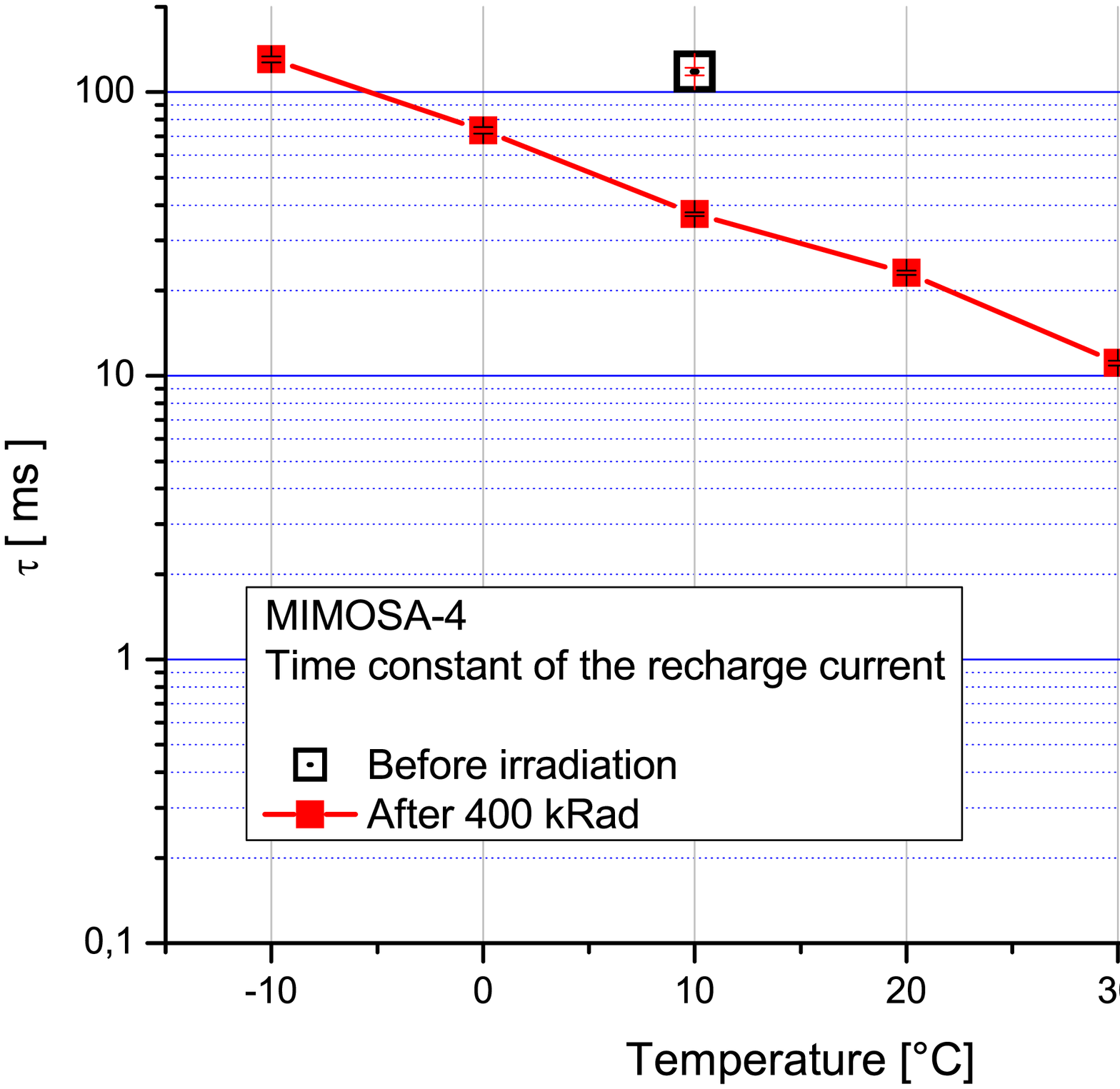}
 \end{minipage}
 \begin{minipage}[c]{6.5cm}  \hspace{0.2cm}
 \includegraphics[width=5.5cm]{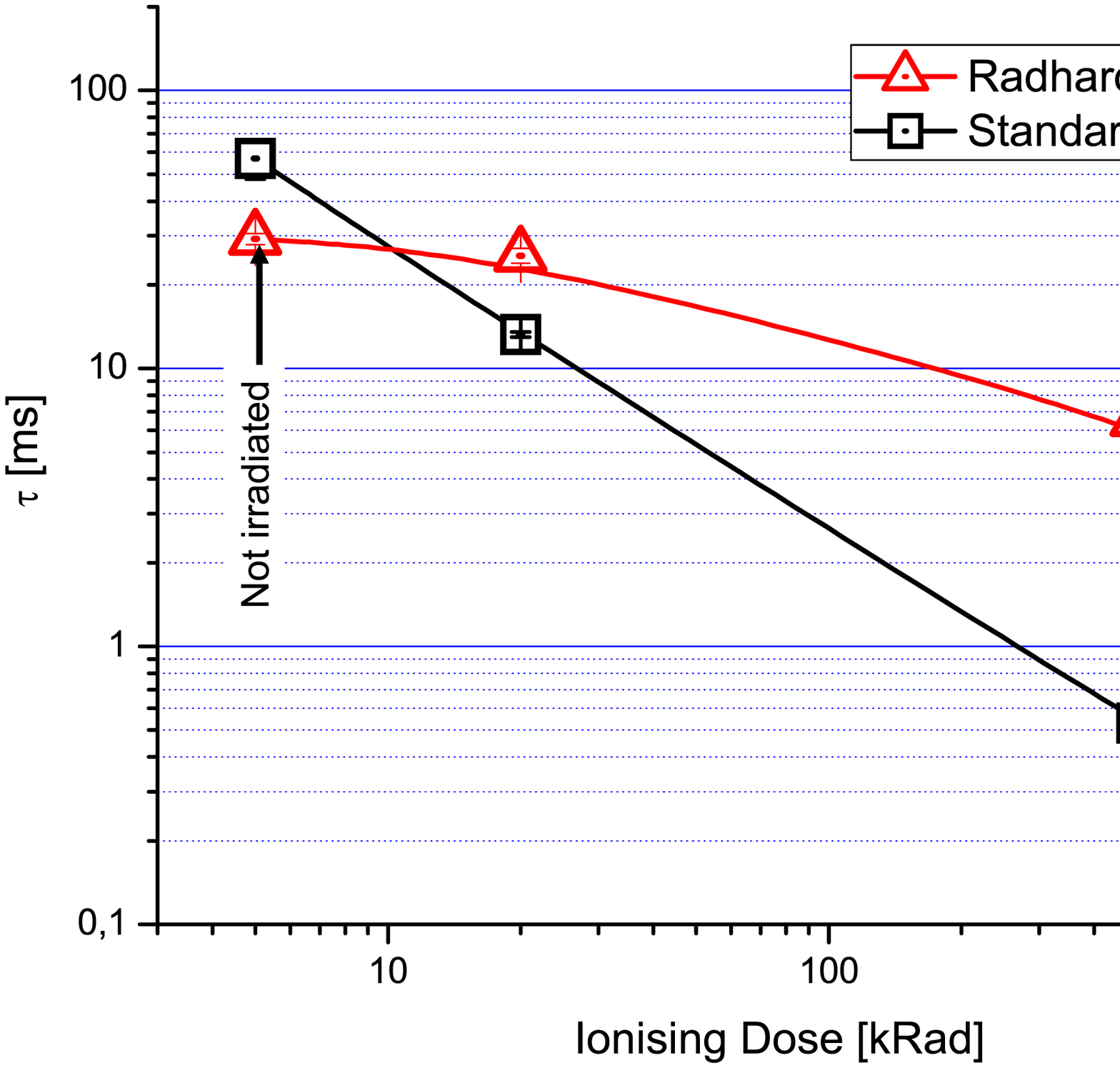}
 \end{minipage}\vspace{1cm}
	\caption[] {Left: The time constant $\tau$ of a SB-pixel as
          function of temperature before and after irradiation. Right:
          The time constant $\tau$ of the standard and the radiation
          hard pixel as function of radiation dose at $T=40
          ~^{\circ}$C. Note that the measurements were done with
          slightly different pixel structures implemented in two
          different processes \mbox{(Left: AMS 0.35 $\mu$m HiRes,}
          Right: AMS 0.35 $\mu$m - OPTO). Colors online.}

\end{center}
  
   \label{Mi11Tau}
\end{figure}

As cooling the sensors is not always feasible, we tried to alleviate
the radiation induced increase of $I_L$ by means of modifying the
collection diode of the sensor. Our study addressed two presumed
sources of radiation induced leakage currents: the thermal generation
of e/h-pairs via radiation induced surface defects at Si/SiO$_2$
interfaces close the collection diode; and the appearance of
conduction paths along this interface due to a radiation induced
positive charge build-up in the SiO$_2$. To suppress those sources, we
surrounded the diode with thin gate SiO$_2$, which is known to show
only few radiation induced defects with respect to the thicker
standard oxide \cite{EnclosedTransistor}. Silicon guard rings were put
to cut the suspected conduction paths, which were assumed to collect
e/h-pairs from regions distant to the diode.

The studies were performed with several generations of chips
implemented in the AMS 0.35 $\mu$m - OPTO process
\mbox{(MIMOSA-9,}\mbox{-11,} and \mbox{-15}). The design of the diode
showing best tolerance to ionizing radiation was inspired by the
design of an enclosed transistor \cite{EnclosedTransistor}, where the
inner transistor ring is formed by the N-well implantation of the
diode. The outer ring is realized with a P-diff implantation, which
serves as guard ring cutting potential conduction paths along the
Si/SiO$_2$ interface of the chip. A gate (set to ground potential) was
placed between both rings to remove any thick SiO$_2$ structures from
the vicinity of the collection diode.

The small ($\sim$ fA) leakage currents of our SB-pixels are not
accessible for direct measurements. The improved diode structure were
thus benchmarked according to the noise and $\tau$ of the pixels.  We
found that previous to irradiation, the radiation hardened pixel shows
slightly higher noise (see figure \ref{Mi11Noise2}) and lower $\tau$
(see figure \ref{Mi11Tau}, right) than the standard pixel. Given the
excellent performances of both designs, this disadvantage is of
marginal relevance and vanishes already after a dose of 10
kRad. Hereafter, the radiation hardened pixel showed lower noise and a
crucially slower time constant. As expected, the improvement in $\tau$
alleviates the drop in gain observed in the response of the chip to
the $^{55}$Fe-source (see figure \ref{Mi11SpectrumA0S2-1MRad},
right). This proves that the modified pixel reaches a higher radiation
hardness that the standard one.

\section{Summary and Conclusion}

Within this work, the effects of ionizing radiation MAPS detectors
with SB-pixels were discussed. We observed two major effects, which
are an increase of shot noise and a drop in the effective gain of the
pixel due to an acceleration of its clearing process. As much as the
shot noise, the latter effect is caused by the known radiation induced
increase of the leakage current of the pixel, which is therefore the
only primary radiation damage observed for X-ray doses up to 1
Mrad. All effects can be alleviated substantially by operating the
detector with moderate cooling and with small integration time.

We studied different diode structures aiming for reducing the leakage
current of irradiated pixels. Best results were achieved by adding two
guard rings, a pseudo-gate and a p-diff implantation, around the
collection diode. Complementary to the cooling, this structure reduced
the shot noise of the pixel and decelerated the clearing process by up
to one order of magnitude, which substantially expands the radiation
hardness of the sensors.

\section*{Acknowledgments}

This work was supported by 
BMBF (06FY1731), GSI (F\&E) and
the Hessian LOEWE initiative through the Helmholtz 
International Center for FAIR (HIC for FAIR).



\end{document}